# Deep DIC: Deep Learning-Based Digital Image Correlation for End-to-End Displacement and Strain Measurement

Ru Yang[1], Yang Li[2], Danielle Zeng[2], and Ping Guo[1,*]

[1] Department of Mechanical Engineering, Northwestern University, Evanston, IL 60208, USA
[2] Research and Advanced Engineering, Ford Motor Company, Dearborn, MI, 48121, USA
* Corresponding author: Ping Guo; ping.guo@northwestern.edu

**Abstract**

Digital image correlation (DIC) has become an industry standard to retrieve accurate displacement and strain measurement in tensile testing and other material characterization. Though traditional DIC offers a high precision estimation of deformation for general tensile testing cases, the prediction becomes unstable at large deformation or when the speckle patterns start to tear. In addition, traditional DIC requires a long computation time and often produces a low spatial resolution output affected by filtering and speckle pattern quality. To address these challenges, we propose a new deep learning-based DIC approach – *Deep DIC*, in which two convolutional neural networks, *DisplacementNet and StrainNet*, are designed to work together for end-to-end prediction of displacements and strains. *DisplacementNet* predicts the displacement field and adaptively tracks a region of interest. *StrainNet* predicts the strain field directly from the image input without relying on the displacement prediction, which significantly improves the strain prediction accuracy. A new dataset generation method is developed to synthesize a realistic and comprehensive dataset, including the generation of speckle patterns and the deformation of the speckle image with synthetic displacement fields. Though trained on synthetic datasets only, *Deep DIC* gives highly consistent and comparable predictions of displacement and strain with those obtained from commercial DIC software for real experiments, while it outperforms commercial software with very robust strain prediction even at large and localized deformation and varied pattern qualities. In addition, *Deep DIC* is capable of real-time prediction of deformation with a calculation time down to milliseconds.

1. Introduction

Digital image correlation (DIC) is a powerful and flexible optical technique that extracts full-field shape, motion, and deformation information through image analysis (Hild and Roux, 2006). It has been widely applied in experimental solid mechanics to accurately measure two-dimensional (2D) and three-dimensional (3D) displacement and strain fields in various material systems (Pan et al., 2009), including engineering metals, polymers (Jerabek et al., 2010), and even bio-materials (Palanca et al., 2016). The common practice of DIC includes pre-painting of a sample with artificial speckle patterns, image capture of sample deformation during material testing, and calculation of full-field displacement and strain fields with a correlation-based algorithm (Sutton et al., 2009). For the calculation of a displacement field, the reference and deformed images are first divided into subsets of the same size. A correlation criterion between all subsets in the reference and deformed images is then established to evaluate their similarity. After matching the reference and deformed subsets, a displacement mapping function uses a sub-pixel registration algorithm to interpolate the matching subsets to sub-pixel displacements. The strain field will then be calculated by taking the spatial derivatives of the obtained displacement field. Compared with pointwise strain gauge measurement (Prabhakaran and Galloway, 2005), DIC is capable of performing a full-field measurement without the need for direct contact with samples, which improves the robustness and accuracy of the measurement. Vison-based DIC also offers an easy-to-set-up solution, which does not require strict experimental conditions, such as a coherent light source and vibration isolation environment, which are often required by interferometric techniques (Dhir and Sikora, 1972).

Since its first introduction in the 1980s, DIC algorithms have been rapidly developed and improved to achieve higher accuracy with better computational efficiency (Sutton et al., 1983). Two fundamental criteria of correlation in DIC, a sum-squared difference (SSD) (Sutton et al., 1983) and cross-correlation (CC) (Sutton et al., 1986), were proposed in the 1980s. Since then, different definitions of correlation criteria have been developed based on the above two fundamental criteria, such as zero-normalized cross-correlation (ZNCC) and a parametric sum of squared difference (PSSD) (Pan, 2011). Besides the correlation criterion, displacement field calculation is another essential step. After finding the image similarity by searching the maximum CC coefficient or the minimum SSD coefficient, a variety of registration algorithms have been developed to derive sub-pixel displacement distributions. Most algorithms can be cast into two categories, the local subset-based (Bruck et al., 1989) and global (continuum) methods (Cheng et al., 2002). Local subset-based methods are realized by interpolation using gray-scale pixel values (Peters and Ranson, 1982) or a correlation matrix (Chen et al., 1993) within each subset. Other subset-based methods are achieved by iterative calculations that solve non-linear mapping parametric vectors (Bruck et al., 1989) and spatial gradients (Pan et al., 2006), or directly finding the local maximum of statistical similarity function (Ronneberger et al., 2015). Since the subset-based methods solve the displacement field within each subset, the calculation can be implemented parallelly to accelerate the overall calculation speed. However, the continuity between different subsets cannot be guaranteed, causing a noisy strain field output. On the other hand, in global (continuum) methods, the displacement field of the whole image is represented by a set of shape functions and solved with finite element methods (Sun et al., 2005). The global (continuum) methods ensure that the whole displacement field is compatible to capture locally heterogeneous deformation, but the overall prediction precision and computational efficiency are inferior to subset-based methods (Wang and Pan, 2016).

As deep learning has received great success in multiple computer vision tasks such as image classification (Krizhevsky et al., 2012), object detection (Ouyang et al., 2016) and 3D reconstruction (Chen et al., 2018), it has also been used in optical flow estimation, a computer vision task that also aims at extracting a displacement field from image pairs (Zayouna et al., 2011). Convolutional neural network (CNN)-based methods have surpassed the traditional optical flow techniques in terms of accuracy and computation speed (Hur and Roth, 2020). By stacking multiple convolutional and deconvolutional layers with proper pooling and activation functions (Ilg et al., 2017), CNN owns a superb ability to recover optical flow fields with sub-pixel accuracy between image pairs, even for large displacement (Sun et al., 2018). By looking into the principle of CNNs, we can find some similarities between CNN and DIC algorithms. The subset correlation calculation in DIC and the convolution operation in CNN are all kernel-based. The peak searching in DIC works similarly to the max-pooling layer in CNN. The difference observed between DIC and CNN is that the correlation criterion in DIC is a highly nonlinear function; while in CNN, feature maps are extracted with a linear calculation of kernel values followed by an activation function. By stacking multiple layers, CNN-based methods are able to recover a highly non-linear relationship between the input and output, potentially outperforming traditional DIC algorithms.

There have been several recent attempts to bring deep learning to DIC. Min et al. (2019) developed a 3D convolutional neural network to extract both the spatial and temporal domain features from a sequence of image sets and output an average displacement vector for each image subset. The training dataset was augmented from a small set of experimental results, which limited their model performance. The strain field prediction was not achieved, while the displacement field prediction was not outperforming traditional DIC. Boukhtache et al. (2021) took the inspiration of deep learning in optical flow and applied it to DIC. They trained multiple CNNs modified from existing optical flow CNNs with synthesized speckle image datasets to achieve high prediction accuracy for sub-pixel deformation or motion. Since their approach targeted sub-pixel displacements, the final displacement field was obtained by first applying a traditional correlation method to retrieve integer shifts followed by a CNN prediction to extract sub-pixel deformation. The approach demonstrated some promising results with high accuracy, but it essentially worked as a hybrid method that still involved subset division, post-filtering, and traditional correlation methods.

Though bringing deep learning to DIC for material characterization seems an attractive and promising idea, there has not been a real success. We have summarized three main challenges that prevent deep learning from being successfully applied in DIC for deformation measurement. (1) No full-field strain field prediction has been reliably demonstrated using deep learning in previous works. The pixel-level prediction enabled by CNNs will inevitably introduce high spatial-frequency noises that will be magnified by the derivative operations in the calculation of the strain field. A Gaussian filter is often applied to smooth the displacement field for strain calculation, but it would defeat the advantage of CNN-based approaches that can potentially capture high spatial-frequency deformation. (2) Previous deep learning-based methods did not show a significant performance advantage over traditional DIC except for computational efficiency. We suspect the reason is partially due to the bad quality of training datasets. Min et al. (2019) generated a training set by augmenting a small set of experimental results with too few variances, which affected the model's transferability and robustness. The ground truth was obtained using traditional DIC, which set its performance limit. In other words, their proposed neural network was designed not to surpass traditional DIC. Boukhtache et al. (2021) applied random displacements at predefined mesh grids and linearly interpolated the displacements inside each cell. In their case, the displacement field was piecewise continuous but not physically informed. The training set would not resemble a typical loading case in actual mechanical tests. (3) There has not been any rigorous attempt to directly compare the prediction accuracy of both displacement and strain fields for deep learning-based and traditional DIC. The real-life performance of deep learning-based DIC is still questionable.

In addition to the challenges mentioned above, the motivation to bring deep learning to DIC has not been very clear in previous studies. DIC is a well-established method with commercially available and industry-trusted software packages, so what are the potential benefits to use deep learning in DIC? In our daily material testing, we found some deficiencies in traditional DIC. When performing a tensile test on soft materials, the magnitude of strain can be well above 100%, where the commercial DIC software will fail to give strain prediction when the speckle patterns start to tear or break, as shown in **Figure 1**. The correlation algorithm is based on pattern matching, which requires the complete integrity of speckle patterns. With the increasing popularity of ultra-stretchable materials, it is imperative to develop new techniques to be able to give robust predictions of full-field strain even when the quality of speckle patterns starts to deteriorate at large deformation. In addition, the computational cost of traditional DIC is still relatively high and significantly affected by the pattern quality.

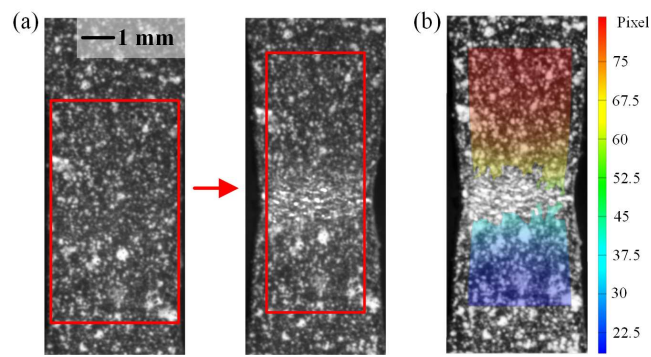

**Figure 1. (a) Speckle patterns start to tear under large deformation and (b) commercial DIC software fails to output displacement prediction in the severely distorted region.**

Motivated by this actual need in our material testing tasks to measure full-field large strain distributions, we develop a new and end-to-end deep learning-based DIC approach – *Deep DIC*, that directly solves the displacement and strain fields from image pairs with no interpolation or iteration. The goal is to achieve robust and accurate predictions of both full-field and high-resolution displacement and strain fields using an end-to-end approach from a sequence of speckle patterns, particularly in tensile testing applications. Furthermore, as inspired by CNN-based optical flow methods, we would like to leverage the ability of CNNs to map highly nonlinear relationships between input and output to overcome the difficulties in estimating large strains with deteriorated speckle patterns. Specifically, it is targeted to be directly compared with commercial DIC software to (1) give a more robust strain prediction at large deformation; (2) achieve a similar or better prediction accuracy for small and moderate deformation; and (3) reduce computing time for potential real-time measurement and prediction.

Facing the same challenges when bringing deep learning to DIC as analyzed above, we propose two major innovations in our approaches to address these challenges. First, rather than calculate the strain fields from the spatial derivatives of the displacement field as traditional DIC does, *Deep DIC* will directly output the strain field from the image input in an end-to-end approach. Two separate CNNs will be designed based on a modified encoder-decoder structure, as referred to as *DisplacementNet* and *StrainNet*. Two CNNs work independently to give displacement and strain field predictions, as well as collaboratively to adaptively update the region of interest (ROI) for tracking large deformation. Second, we design a new method to synthesize realistic and comprehensive datasets for training our model. By rendering speckle patterns with different qualities, as well as prescribing a wide variety of random rigid body motion and deformation, we can increase the robustness and adaptability of *Deep DIC*. Though only trained on synthesized datasets (which could be a potential benefit with a very low training cost), *Deep DIC* is able to outperform traditional DIC on real experimental data. In addition to these two innovations, we also systematically evaluate the performance of *Deep DIC* and compare it with commercial DIC software to validate its real-life performance. The insights gained from the performance analyses will also provide scientific and archival values to the research community.

## 2. Methods

The overall workflow of *Deep DIC* for tensile testing is illustrated in **Figure 2**. A region of interest (ROI) is initially defined in the starting frame of a sequence of image inputs. Two individual CNNs, *DisplacementNet* and *StrainNet*, are designed to separately calculate the displacement and strain fields in the ROI extracted from the image inputs. The accumulated displacement and strain fields are updated based on the latest incremental calculations. The definition of ROI is then updated based on the updated coordinates of four corner points in the accumulated displacement field, so the ROI is adaptively changed to track large deformation. The procedure is repeated to analyze the next pair of image inputs with an updated ROI until the last frame in the sequence. The two CNNs provide independent predictions of displacement and strain fields in an end-to-end manner directly from raw image inputs.

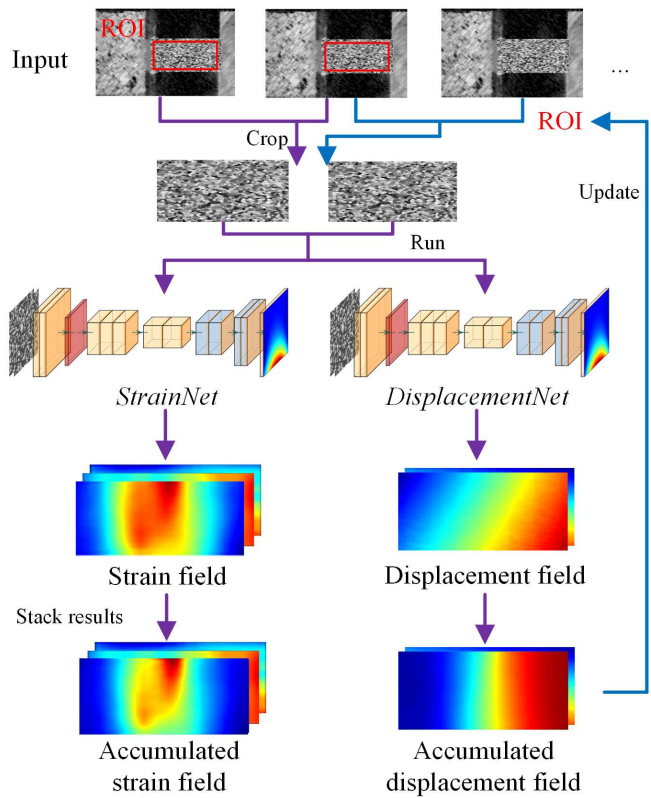

**Figure 2. Workflow of *Deep DIC*.**

In the following sections, we introduce the designs of *DisplacementNet* and *StrainNet*, the methodology to generate synthetic training datasets and corresponding ground truths, as well as training details.

## 2.1. *Deep DIC* architecture

We design two separate CNNs, *DisplacementNet* and *StrainNet*, to independently learn the displacement and strain fields from the same input of an image pair. The schematics of their architectures are illustrated in **Figure 3**.

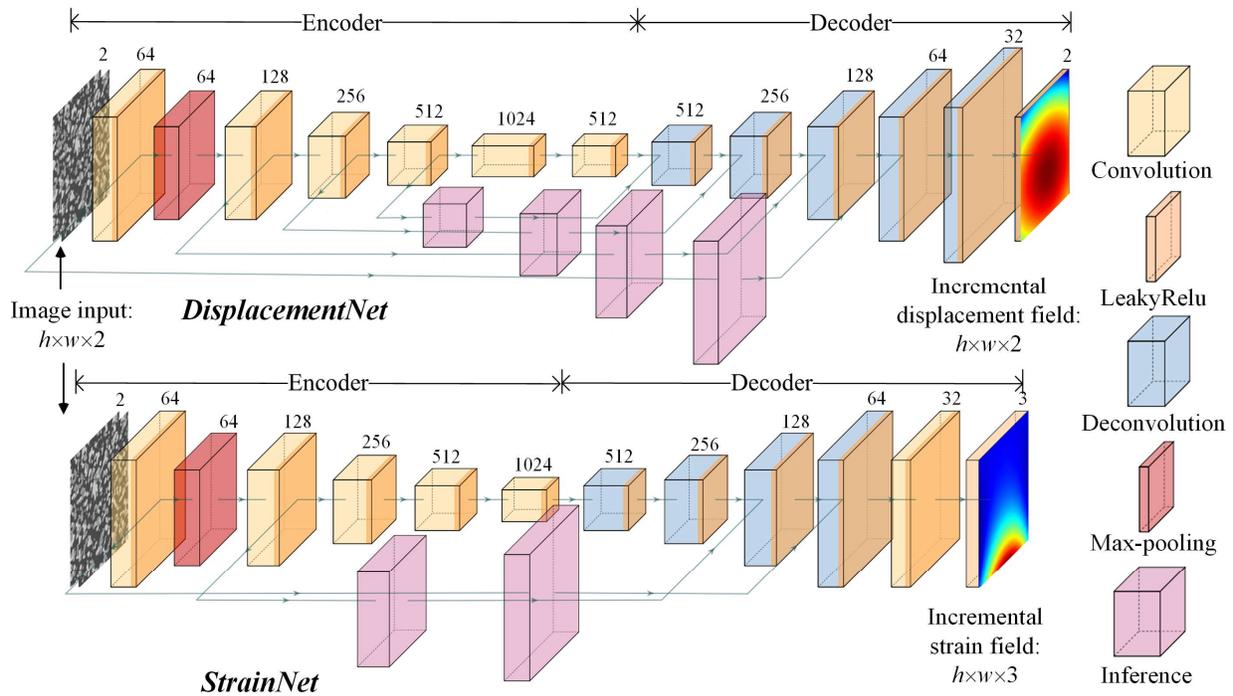

**Figure 3. CNN architecture of *DisplacementNet* and *StrainNet*. The numbers above each module indicate the feature map depth.**

The input to both models is a pair of speckle images with a height $h$ and width $w$. Due to the adaptive tracking of ROI, the input image size is not fixed. Pre-processing is needed to scale the image to the nearest multiples of 32 in both length and width, so that the exact size matching can be guaranteed for the inference concatenation. The results are then scaled back to get the actual displacement field. The strain field prediction is not affected by the rescaling. The final outputs from *DisplacementNet* are two images of size $h \times w$, giving the predicted displacement components, $u$ and $v$. The outputs from *StrainNet* are three images of size $h \times w$, giving the three plane strain components, $\varepsilon_{xx}$, $\varepsilon_{yy}$, and $\varepsilon_{xy}$. In our training dataset, all the inputs are formatted to the size of $128 \times 128$, which are detailed in the next section.

Both *DisplacementNet* and *StrainNet* follow a modified encoder-decoder structure, which has been widely adopted in image segmentation tasks that require high-resolution output (Badrinarayanan et al., 2017). In the encoder part, a chain of convolution operations with a kernel size of 3 and a stride size of 2 sequentially condenses the size of the feature map while doubling its depth with each convolutional layer. This allows the CNNs to extract deep features from the sparse information in the input image pair. In the decoder part, a chain of deconvolution operations reverses the encoder operations to double the feature map size and halve the map depth with each deconvolutional layer. The function of the deconvolutional layer is to recover the high-resolution displacement/strain field from high-dimensional feature maps. Since the absolute values for strain and displacement are numerically small, the gradient of the loss function with respect to the CNN parameters could vanish as the network goes deep. Therefore, in order to accelerate training, for each convolutional (deconvolutional) layer, we use a batch normalization operation before the activation function (Ioffe and Szegedy, 2015). In both CNNs, following each batch normalization operation, we adopt LeakyReLU (Maas et al., 2013) as the activation function with a slope of 0.01 for negative values.

Inspired by DenseNet (Huang et al., 2017), we modify the encoder-decoder structure by adding multiple inference layers to concatenate early-stage feature maps in the encoder stage to features maps in the decoder stage. This operation is intended to prevent the loss of details in the chain convolution operations. We find the inclusion of inference layers improves the training speed and prediction accuracy. It is noted that *DisplacementNet* and *StrainNet* have slightly different structures in terms of the depth and number of inference layers, which have been manually adjusted to achieve the best learning results.

### 2.2. Dataset generation

We propose to train *Deep DIC* completely on synthetic datasets. This allows significant cost savings and provides better control over data quality. In this section, we describe the method to generate a realistic and high-quality dataset with both reference and deformed images as well as the corresponding ground truths of displacement and strain fields. The overall approach for dataset generation is shown in **Figure 4.** We first create artificially generated speckle pattern images with different qualities in order to increase the robustness of trained models to deal with real-life situations. A variety of random motion and deformation is analytically defined to generate a displacement field as the ground truth for *DisplacementNet*. Based on the defined displacement field, the original speckle pattern image is warped to get the deformed image. The ground truth for the strain field can be analytically calculated by taking the spatial derivatives of the displacement field. Additional post-processing, such as random crop, down-sampling, and adding artificial noises, etc., is performed to get a complete set of one data sample, which includes the inputs to *Deep DIC*: the reference and deformed images, and the outputs: the predefined displacement field and calculated strain field. The implementation details are described step by step as follows.

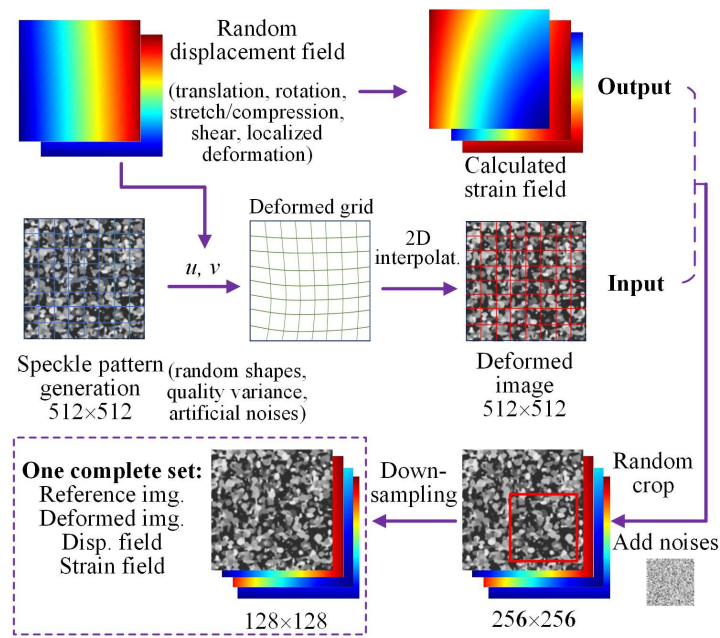

**Figure 4. Schematic of the dataset generation workflow.**

**Speckle pattern generation**

Speckle pattern images are generated by stacking ellipses with random sizes and gray-scale values. Each speckle pattern image contains 2,800 to 4,500 ellipses within a frame size of 512 × 512. For each sample in the dataset, a unique and random speckle pattern is created, so there is no re-utilization of speckle images. To increase the robustness and adaptivity of *Deep DIC*, we deliberately include speckle patterns with quality variances, including images with sparse speckle distribution (5% of the total samples), random large speckles (30%), extra noises (5%), and low contrast (5%). The detailed speckle pattern generation algorithm and parameter range are given in **Table 1**. Examples of different pattern qualities are demonstrated in **Figure 5**.

**Table 1. Speckle pattern generation algorithm.**

```
Output: Image I₅₁₂ₓ₅₁₂ with randomly generated speckle patterns

   1. Define a random density       n∈[2,800, 4,500]
      number
   2. Define random ellipse         Pₙₓ₂ ← (xᵢ,yᵢ), i=1,2,…,n
      center positions              xᵢ,yᵢ∈[0, 511]
   3. Define random long and        Bₙₓ₂ ← (lᵢ,sᵢ), i=1,2,…,n
      short axis lengths            lᵢ,sᵢ∈[1.2, 6.8]
   4. Define random gray            Gₙₓ₁ ← gᵢ, i=1,2,…,n
      scale values                  gᵢ∈[0.08, 0.98]
   5. for j ← 1 to n do
         I[x,y] ← gⱼ, (x-P[j,1])²/B[j,1]²+(y-P[j,2])²/B[j,2]²<1
      end for
   6. Apply Gaussian blur to I₅₁₂ₓ₅₁₂.
```

```
Sparse pattern:          lᵢ,sᵢ∈[1.2, 3.5]
Large speckles:          Add 1~5 random ellipses with long and short
                         axis lengths between [6.5,9.5] after steps
                         1-5
Extra noise:             Add Gaussian noises with an intensity
                         [0.001-0.01]
Low contrast:            gᵢ∈[0.08, 0.68]
```

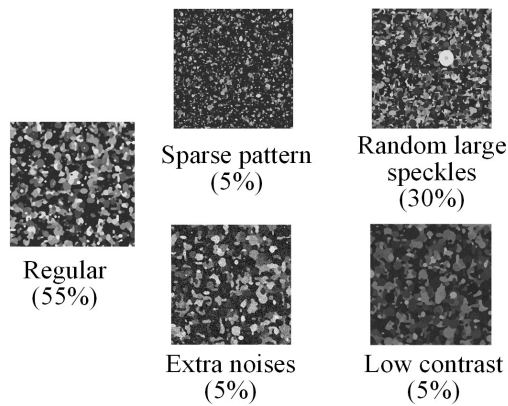

Regular (55%)  
Sparse pattern (5%)  
Random large speckles (30%)  
Extra noises (5%)  
Low contrast (5%)

**Figure 5. Variations of speckle pattern quality in the dataset.**

**Displacement and strain field generation**

A 2D displacement field is defined for each sample image by combining random rigid body translation, rotation, stretch/compression, shear, and localized deformation formulated with 2D Gaussian functions. The mathematical definition of a randomly generalized displacement field is given in Eq. (1), while the localized deformation is described by 2D Gaussian functions in Eq. (2).

$$\begin{bmatrix} u \\ v \end{bmatrix} = \underbrace{\begin{bmatrix} \cos\theta & \sin\theta \\ -\sin\theta & \cos\theta \end{bmatrix}}_{Rigid\ body\ rotation} \cdot \left( \underbrace{\begin{bmatrix} k_x - 1 & \gamma_x \\ \gamma_y & k_y - 1 \end{bmatrix}}_{\substack{Uniform\ stretch \\ and\ shear}} \cdot \begin{bmatrix} x \\ y \end{bmatrix} + \underbrace{\begin{bmatrix} u_x^g \\ u_y^g \end{bmatrix}}_{\substack{2D\ Gaussian \\ deformation}} \right) + \underbrace{\begin{bmatrix} t_x \\ t_y \end{bmatrix}}_{\substack{Rigid\ body \\ translation}} \quad (1)$$

$$\begin{bmatrix} u_x^g \\ u_y^g \end{bmatrix} = \sum_{j=1}^{N} \begin{bmatrix} A_x^j e^{-\frac{1}{2}\left(\frac{x-x_0^j}{\sigma_{x_0}^j}\right)^2 - \frac{1}{2}\left(\frac{y-y_0^j}{\sigma_{y_0}^j}\right)^2} \\ A_y^j e^{-\frac{1}{2}\left(\frac{y-y_1^j}{\sigma_{y_1}^j}\right)^2 - \frac{1}{2}\left(\frac{x-x_1^j}{\sigma_{x_1}^j}\right)^2} \end{bmatrix}, \quad N = 1\ or\ 2 \quad (2)$$

{$u$, $v$} are the displacement components in the $x$ and $y$ directions. {$x$, $y$} are the original coordinates of each pixel in the reference image. The range of rigid body translation ($t_x$, $t_y$), rotation ($\theta$), stretch/compression ($k_x$, $k_y$), and shear ($\gamma_x$, $\gamma_y$) are given in **Table 2**. The rotation center is assumed to be at the coordinate origin. The raw image (512 × 512) is later randomly cropped to a size of 256 × 256 to effectively shift the rotation center to a random position in the image. Two 2D Gaussian functions, given in Eq. (2), define two localized deformations with randomized amplitudes ($A_x$, $A_y$), centers of the peak ($x_0$, $y_0$, $x_1$, $y_1$), and standard deviations ($\sigma_{x0}$, $\sigma_{y0}$, $\sigma_{x1}$, $\sigma_{y1}$). Up to two Gaussian function-defined displacement fields might be superimposed to the final displacement field. The range of Gaussian function parameters is also given in **Table 2**. Since the defined image is later randomly cropped from 512 × 512 to 256 × 256, the centers of the peak in the Gaussian functions can be effectively outside of the image frame to add more deformation variances and to remove hidden trends in the dataset.

**Table 2. Definition and range of displacement parameters.**

| Deformation parameter | Range | Effective max. displacement (w.r.t the final image size 128×128) |
|---|---|---|
| *Translation* | | |
| $t_x, t_y$ | -4.0 ~ 4.0 pixel | 2.0 pixel |
| *Stretch* | | |
| $k_x, k_y$ | 0.96 to 1.04 | 5.1 pixel |
| *Rotation* | | |
| $\theta$ | -0.01 to 0.01 rad | 2.4 pixel |
| *Shear* | | |
| $\gamma_x, \gamma_y$ | -0.03 to 0.03 | 3.8 pixel |
| *Gaussian functions* | | |
| $A_x, A_y$ | 0.003 ~ 0.6 | 4.6 pixel |
| $\sigma_{x0}, \sigma_{y0}, \sigma_{x1}, \sigma_{y1}$ | 0.06 ~ 0.5 | |
| $x_0, y_0, x_1, y_1$ | 0 ~ 511 | |

The generated 2D displacement field is adopted as the ground truth for training *DisplacementNet*. The corresponding strain field can be analytically calculated by taking the spatial derivatives of the displacement field based on the infinitesimal strain assumption, which is defined in Eq. (3). The calculated strain field is used as the ground truth for training *StrainNet*. Since the random displacement fields are defined by Gaussian functions, which are smooth or indefinitely differentiable, the compatibility of corresponding strain fields is always satisfied.

$$\begin{aligned}
\varepsilon_{xx} &= k_x + \frac{\partial u_x^g}{\partial x} \\
\varepsilon_{yy} &= k_y + \frac{\partial u_y^g}{\partial y} \\
\varepsilon_{xy} &= \frac{1}{2}\left(\gamma_x + \gamma_y + \frac{\partial u_x^g}{\partial y} + \frac{\partial u_y^g}{\partial x}\right)
\end{aligned} \quad (3)$$

**Image deformation**

The deformed image is synthesized by first applying the predefined displacements to each pixel to get the deformed grid coordinates and then interpolating the randomly scattered grids back to a uniform grid using MATLAB in-built function *griddata*. We randomly crop the reference and warped images from 512

× 512 to a size of 256 × 256 to remove hidden patterns in the dataset. Additional Gaussian noises with an intensity of 0.001 and a mean value of 0 are applied separately to the reference and warped images to mimic the image capture noises. The images are further downsampled to 128 × 128 to blur the sharp edges. Two examples of generated displacement and strain fields, along with the reference and deformed speckle images, are demonstrated in **Figure 6**. The statistical analysis of the dataset can be referred to in detail in Appendix A.

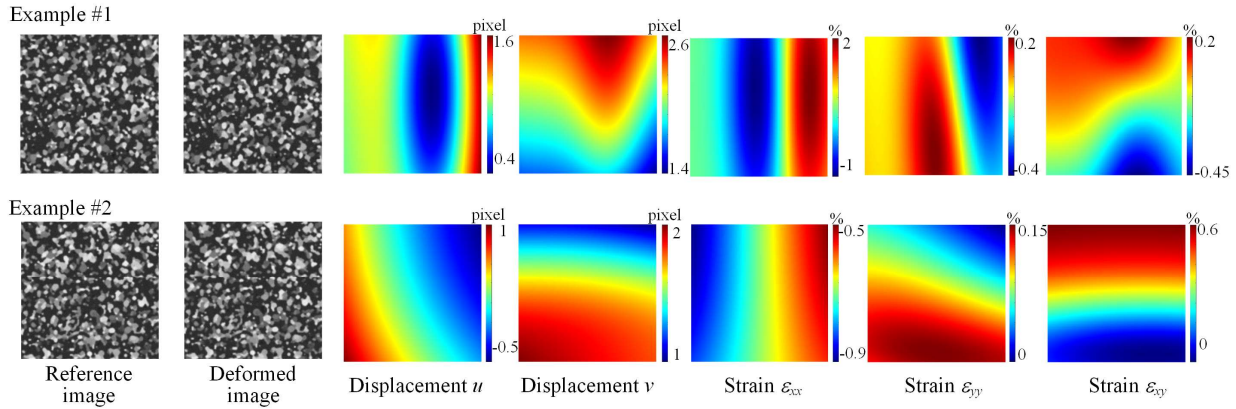

**Figure 6. Two data samples in the training set with reference and deformed speckle images, corresponding displacement fields, and calculated strain fields.**

### 2.3. Training details[1]

*DisplacementNet* and *StrainNet* are both implemented on the PyTorch (version 1.6.0) platform (Paszke et al., 2019). The package Torchvision (version 0.7.0) is used to build the CNN structure and Pillow (7.2.0) is used to load, crop and resize the images. The loss function for *DisplacementNet* is the mean square error (MSE) between the predicted and predefined displacement fields multiplied by 10. The loss function for *StrainNet* is the MSE between the predicted and ground truth strain fields multiplied by 100 to compensate for the scale of strain values. We choose Adam (Kingma and Ba, 2015) as the optimization method since it can adaptively change the learning rate according to the current gradient, resulting in a faster convergence rate. The two momentum parameters for Adam are set to $\beta_1 = 0.9$ and $\beta_2 = 0.999$, as recommended by

---

[1] The code, model, and dataset are released on GitHub: https://github.com/RuYangNU/Deep-Dic-deep-learning-based-digital-image-correlation

Kingma and Ba (2015). For *DisplacementNet*, the learning rate is initiated with 0.001 and further reduced by a factor of 100 after 100 epochs. After 200 epochs of training, the error in the validation set for *DisplacementNet* is settled below 0.01. For *StrainNet*, the learning rate starts at 0.001 and is reduced to 1e-5 after 100 epochs. The training is stopped at epoch 198 for *StrainNet* when the validation error is settled to 0.06. The convergence history of both *DisplacementNet* and *StrainNet* is plotted in **Figure 7.** The mean maximum prediction errors and the average errors on the validation set are summarized in **Table 3**. Since strain is represented in percentage, the strain error indicated in this table and the whole paper is the absolute value as a percent strain, not the relative percentage error.

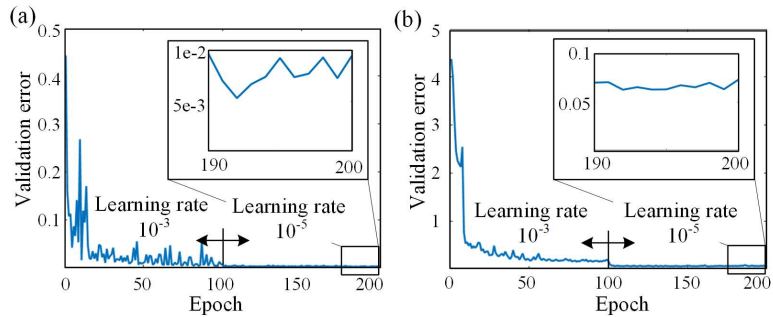

Figure 7. Convergence history of (a) *DisplacementNet* and (b) *StrainNet*.

Table 3. Performance of *Deep DIC* on the validation and test sets.

| *Prediction error* | **DisplacementNet (pixel)** | | **StrainNet (%)** | |
|---|---|---|---|---|
| *Validation set* | Max: 0.047 | Average: 0.024 | Max: 0.064 | Average: 0.031 |
| *Test set* | Max: 0.083 | Average: 0.038 | Max: 0.085 | Average: 0.041 |

\* Strain error is represented as the absolute value in terms of percent strain, not a relative percentage error.

3. **Results and Discussion**

*Deep DIC* is only trained on a synthetic dataset, but is designed to perform on both simulated and experimental data. In this section, we first provide a discussion on the adoption of an end-to-end approach for strain prediction. Then we systematically evaluate the performance of *Deep DIC* on both synthetic samples and experimental data. The results are directly compared with commercial DIC software, VIC-2D (v6, *Correlated Solutions, Inc.*, USA) (Correlated Solutions, 2021) and GOM Correlate (v2020, *GOM Metrology*, Germany) (GOM, 2021). In addition to the comparison of predicted displacement fields, we specifically include the result comparison of strain field prediction for its important application in material testing.

### 3.1. Discussions on neural network architecture

**<u>Direct strain prediction from *StrainNet*</u>**

One major difference between the *Deep DIC* and previous attempts is the direct prediction of a strain field from a pair of image inputs, independent of displacement predictions. We have noticed significant advantages of this end-to-end method over the approach to take spatial derivatives with respect to the displacement field. Even in traditional DIC, spatial filtering is commonly adopted to compute the strain field, which not only reduces the spatial resolution of the strain prediction (Stinville et al., 2016), but also adds another knob tuning parameter in the post-processing, since there is no established guideline on the correct choice of filtering parameters.

The situation gets worse with deep learning-based approaches. The *Deep DIC* and other deep learning-based approaches perform a pixel-wise prediction. Though they can improve the spatial resolution of the prediction as demonstrated by Boukhtache et al. (2021), the predicted displacement field is not guaranteed to be continuous. The analytical derivation of the strain field from the predicted displacement field will enlarge these high-frequency noises which are hard to remove by simple filtering. An accurate prediction of displacements may still lead to large errors and high-frequency noises in the strain prediction if directly calculated from spatial derivatives. As demonstrated by the example shown in **Figure 8**, the maximum error in the displacement field estimation is only 0.016 pixels, while the derived strain field from taking the spatial derivatives shows high-frequency noises and large errors due to the discontinuity of the displacement field. The strain prediction directly from *StrainNet* for the same case is plotted for comparison. The color bars in the error maps are capped at 0.5% for easy visual comparison. The maximum and average prediction errors are 5.93% and 0.24% for the calculated strain field, but only 0.11% and 0.018% for the *StrainNet* prediction.

Besides the better strain prediction accuracy and resolution, the adoption of *StrainNet* also brings the additional benefits of better handling of rigid body rotation. In the calculation of strain fields for the groud truth in the dataset generation, the rigid body rotation is removed from the displacement field, as described in the previous section. The additional rigid body rotation does not affect the associated strain field in the ground truth. Since *StrainNet* directly predicts strains by learning from the given training dataset, it inherits the ability to remove the influence of rotational motion in the strain calculation implicitly through the deep neural networks.

On the flip side, there are more than one strain measures depending on applications. The *StrainNet* is built on an infinitesimal strain assumption given that the deformation between image frames is small. It is not able to output other types of strain measures directly. One possible workaround is to define separate *StrainNets* for each common strain measure, as long as the ground truth can be properly defined according to the specific strain definition.

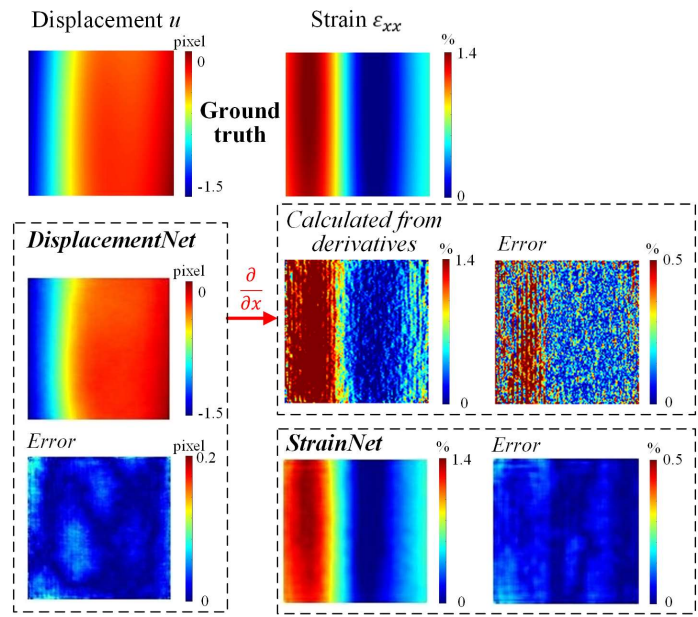

**Figure 8. Comparison of strain field predictions from *StrainNet* and spatial derivates.**

## 3.2. Performance on synthetic images

We first use a test set (150 samples) to compare the predicted results from *DisplacementNet* and *StrainNet* to the ground truth. The performance on the test set is summarized in **Table 3**. The corresponding mean maximum and average displacement errors are 0.083 pixels and 0.038 pixels, while the mean maximum and average strain errors are 0.085% and 0.041%. The test set results are quite impressive and not too far from the accuracy obtained on the validation set. We pick two test examples corresponding to relatively small and large deformation respectively and compare the predicted displacement and strain fields with the ground truth and results obtained from commercial DIC software, VIC-2D. Both examples are run in VIC-2D with a subset size of 7 and a step size of 2. We calculate the average prediction errors for the two displacement components and three strain components for *Deep DIC*. Since VIC-2D has an image output size smaller than the original image input, we interpolate its results to 128 × 128 to match the ground truth image size by MATLAB in-built function *interp2* and then compare them to the ground truth. The results are summarized and compared in **Table 4**.

**Table 4. Performance comparison between *Deep DIC* and VIC-2D on two examples in the test set.**

|  | Average displacement error | | Average strain error | | |
| --- | --- | --- | --- | --- | --- |
| *Test #1* | $u$ | $v$ | $\varepsilon_{xx}$ | $\varepsilon_{yy}$ | $\varepsilon_{xy}$ |
| **Deep DIC** | 0.0079 pixel | 0.0052 pixel | 0.020% | 0.019% | 0.040% |
| *VIC-2D* | 0.0342 pixel | 0.0373 pixel | 0.064% | 0.271% | 0.078% |
| *Test #2* | $u$ | $v$ | $\varepsilon_{xx}$ | $\varepsilon_{yy}$ | $\varepsilon_{xy}$ |
| **Deep DIC** | 0.0097 pixel | 0.0081 pixel | 0.035% | 0.015% | 0.037% |
| *VIC-2D* | 0.067 pixel | 0.0448 pixel | 0.112% | 0.084% | 0.089% |

**Figure 9** shows the result comparison of the first test example with relatively small but complex deformation. *Deep DIC* outputs more accurate predictions for all the displacement and strain components. The average prediction errors from VIC-2D are 2-7 folds of those from *Deep DIC*. The strain prediction errors from VIC-2D are often of the same magnitude or even larger than the predicted strains. Particularly, the strain component $\varepsilon_{yy}$ shows more than a 10-fold difference in terms of average prediction error between the two methods. The prediction accuracy drops for the complex shear strain component $\varepsilon_{xy}$ for both methods, but *Deep DIC* still performs much better to capture the strain pattern. **Figure 10** shows the comparison of the second test example with relatively large and simple deformation. In this case, the shear strain is almost zero, while the sample undergoes non-uniform bilateral stretching. The accuracy comparison indicates a similar conclusion that *Deep DIC* performs much better on the test samples compared with the commercial DIC software.

The major reason for the poor performance of VIC-2D on these two test examples is due to the additional artificial white noises added to the image inputs. Good performance of *Deep DIC* on the test set is expected, since the test data are generated following the same algorithms to generate the training and validation sets (though with different random values). The addition of white noises is well handled by *Deep DIC*, since the CNNs implicitly learn the denoising operation in the deep neural networks.

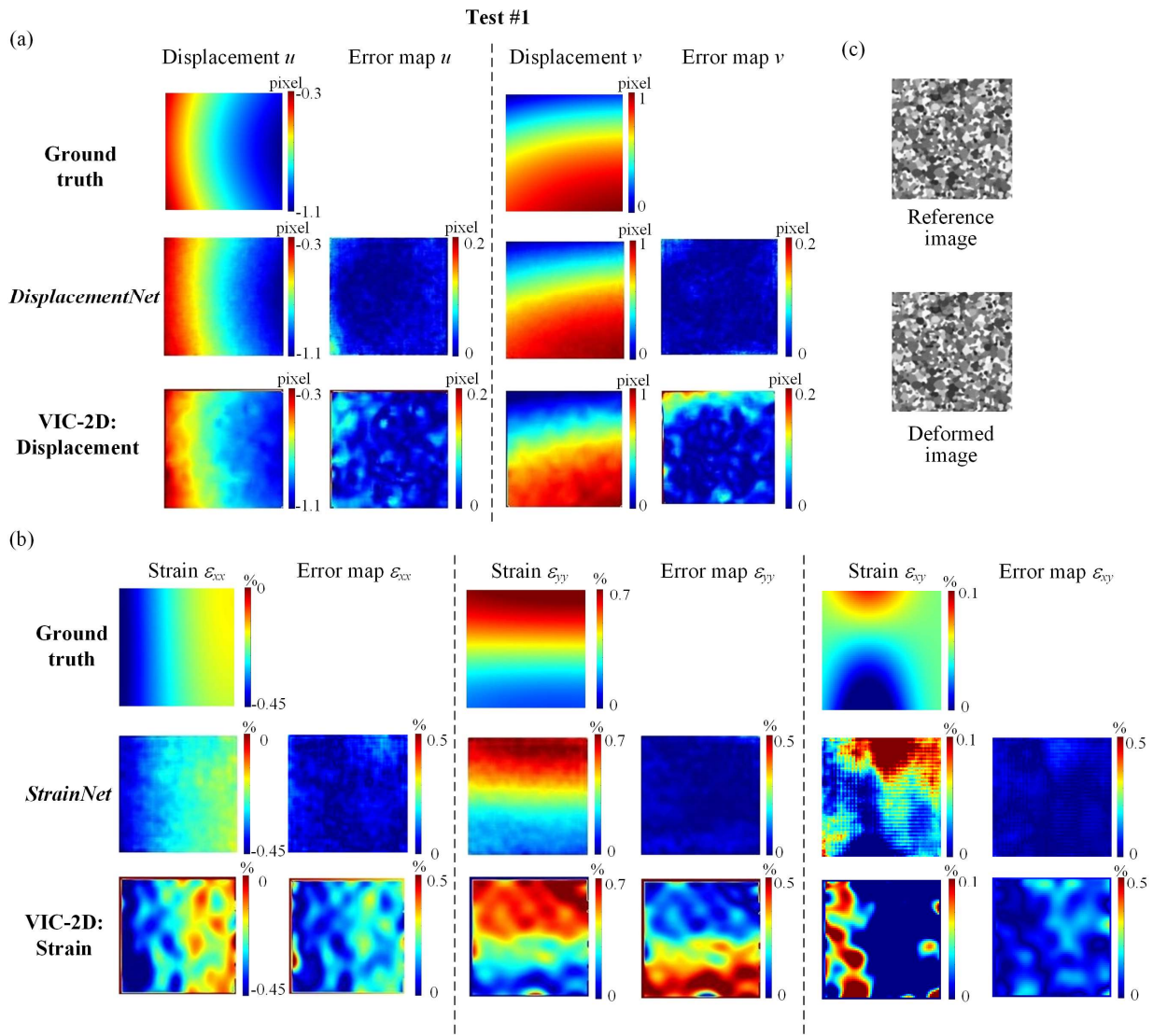

**Figure 9.** Prediction comparison between *Deep DIC* and VIC-2D on one sample in the test dataset with relatively small but complex deformation (Test #1): (a) displacement field; (b) strain field; and (c) input image pair.

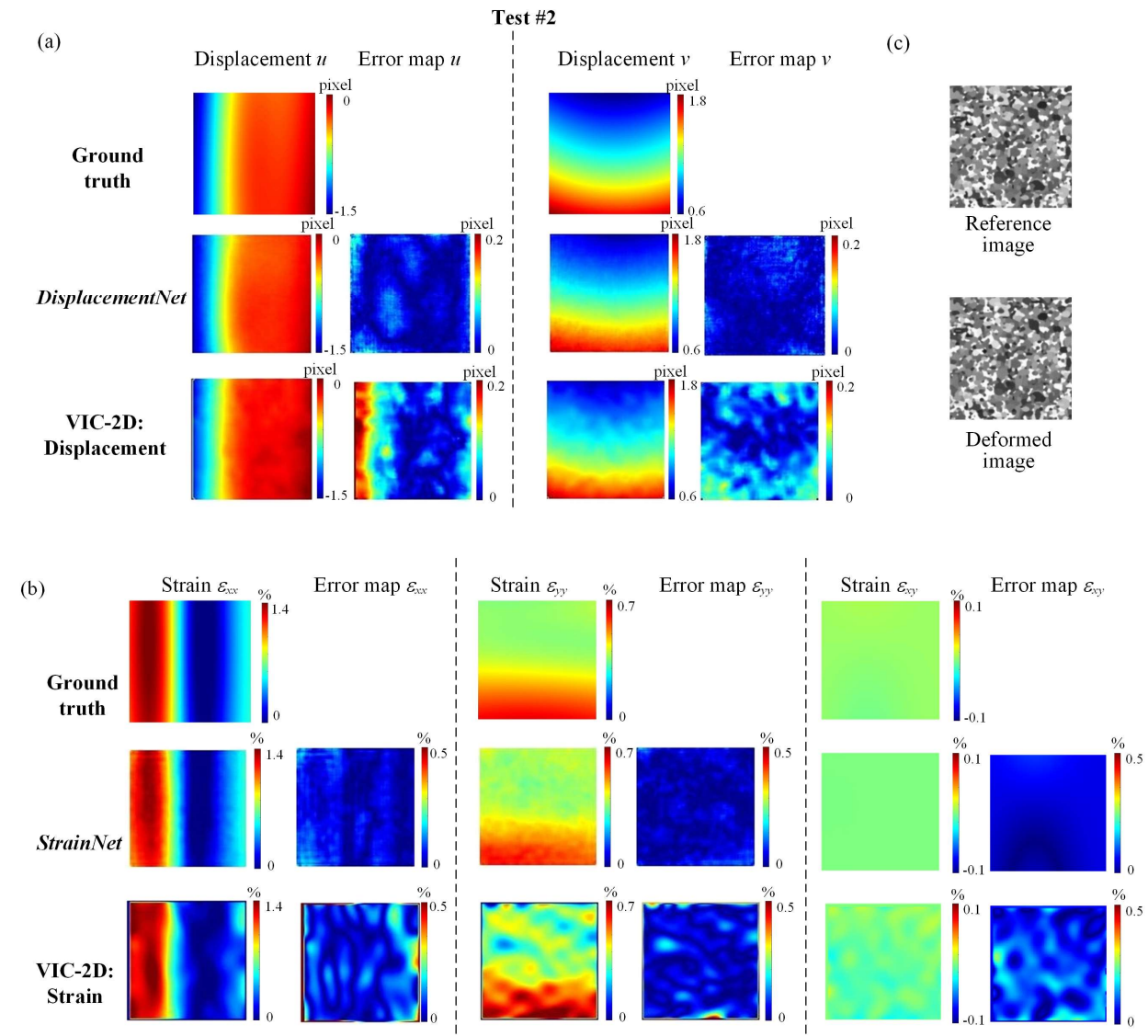

**Figure 10.** Prediction comparison between *Deep DIC* and VIC-2D on one sample in the test dataset with relatively large and simple deformation (Test #2): (a) displacement field; (b) strain field; and (c) input image pair.

### 3.3. Experimental validation

*Deep DIC* uses an end-to-end learning approach, so there is no physically informed knowledge embedded in the system. The only control we have is over how we can design a realistic and comprehensive dataset, so the model can learn to perform the correlation, interpolation, and derivative operations to extract accurate displacement and strain fields. The complexity of the displacement and strain fields may or may not affect the prediction accuracy. In other words, deep learning-based DIC may perform well on a particularly complex case, but perform poorly on a simple scenario, such as stationary image inputs and simple rigid body motion. In this section, we systematically evaluate the noise floor level, rigid body motion prediction, and the real-life performance of displacement and strain predictions in tensile tests. The results are directly compared with commercial DIC software.

**<u>Noise floor level</u>**

We first experimentally evaluate the noise floor of *Deep DIC* and compare the results with those obtained from VIC-2D. 21 pairs of stationary speckle images were captured using a CMOS camera (MQ022MG-CM, *Ximea*, Germany) and a telecentric lens with a fixed working distance of 139 mm and a magnification ratio of 0.3X (#58-428, *Edmund Optics*, USA). The image pairs were fed to both *DisplacementNet* and *StrainNet*. One predicted example with the probability density distributions is plotted in **Figure 11**. The theoretical outputs of displacement and strain fields should be zero and uniform. *Deep DIC* performs better in terms of the noise floor of displacement prediction as compared with VIC-2D. VIC 2D shows a slightly smaller noise floor in the strain field prediction. The statistical analysis of all 21 measurements is summarized in **Table 5**. *DisplacementNet* has an average noise floor of 0.0094 pixels (with a standard deviation of 0.065 pixels) in the displacement prediction, which is slightly better than the commercial DIC software. *StrainNet* has an average noise floor of 0.0073% strain (with a standard deviation of 0.0045%), which is on par with the commercial solution. It should be noted that usually the noise floor level is not evaluated for the strain prediction, since it is derived from the displacement field; however, since the strain prediction in *Deep DIC* is independent of the displacement prediction, it is meaningful to evaluate its noise floor level.

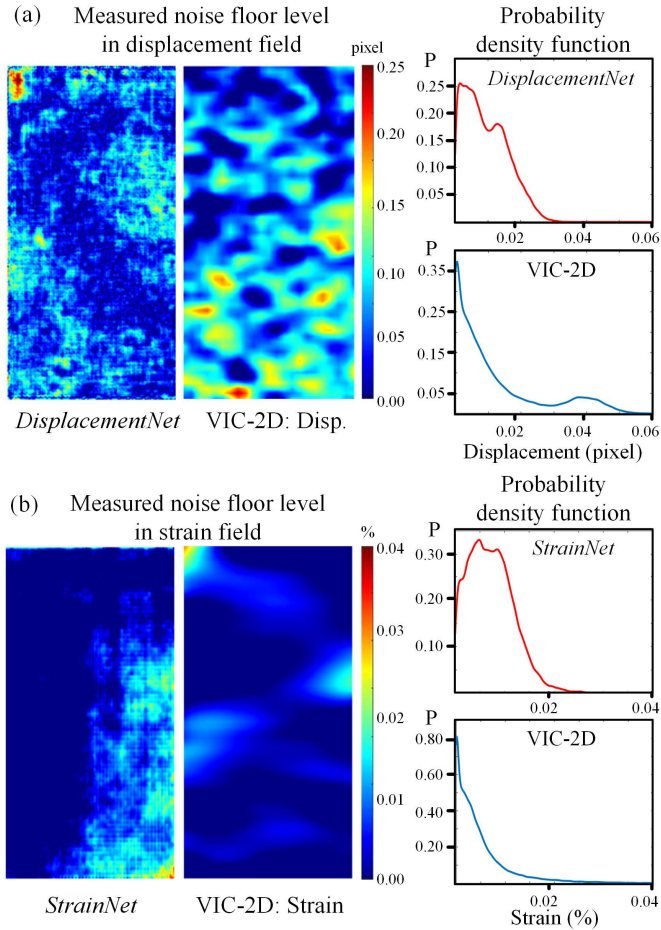

**Figure 11.** Comparison of noise floor measurement of *Deep DIC* and VIC-2D from experimentally captured stationary speckle images: (a) displacement field and (b) strain field. The ideal displacement and strain fields should be uniform and zero.

**Table 5.** Comparison of noise floor level measurement between *Deep DIC* and VIC-2D.

|  |  | Mean value | Standard deviation |
|---|---|---|---|
| *Displacement* | DisplacementNet | 0.0094 pixel | 0.0065 pixel |
|  | VIC-2D | 0.0118 pixel | 0.0134 pixel |
| *Strain* | StrainNet | 0.0073 % | 0.0045 % |
|  | VIC-2D | 0.0052 % | 0.0068 % |

**Validation of rigid body translation**

We validate the *Deep DIC* performance on simple rigid body translational motion. A sample with speckle patterns was clamped only on the moving side of a miniature universal material test system (μts, *Psylotech Inc.*, USA), which has a 25 nm displacement resolution. 19 step motions with a step size of 35 μm in the vertical direction were commanded to move the sample without stretching it. The same camera and lens system was adopted from the noise floor measurement to capture the sample image after each step motion. A total of 20 images including the starting position were analyzed. The predicted final displacement field is plotted in **Figure 12(a)**. The ideal output should be a uniform field. By averaging the whole displacement field, we take the average value as the predicted translational motion. We compare the predicted translational motion from *DisplacementNet* with VID-2D in **Figure 12(b)**. The error bar in the figure indicates the maximum difference within the predicted displacement field for each frame. The difference between *DisplacementNet* and VIC-2D is plotted as the error curve in **Figure 12(b)**. The maximum difference is 0.275 pixels in the 8th frame.

*DisplacementNet* performs less impressive in this test. It produces non-uniform displacement predictions and noticeable differences from VIC-2D results. This is largely attributed to the lack of pure rigid body motion samples in our training dataset. Based on how we define the random displacement fields according to Eqs. (1) and (2), no uniform displacement field is included in the dataset. By adding additional data samples with pure translation and/or rotation will help to improve the performance of *DisplacementNet* in predicting rigid body motion. On the other hand, the rather mediocre performance of *Deep DIC* in this task does not indicate its ability to handle complex deformation situations due to the nature of deep learning.

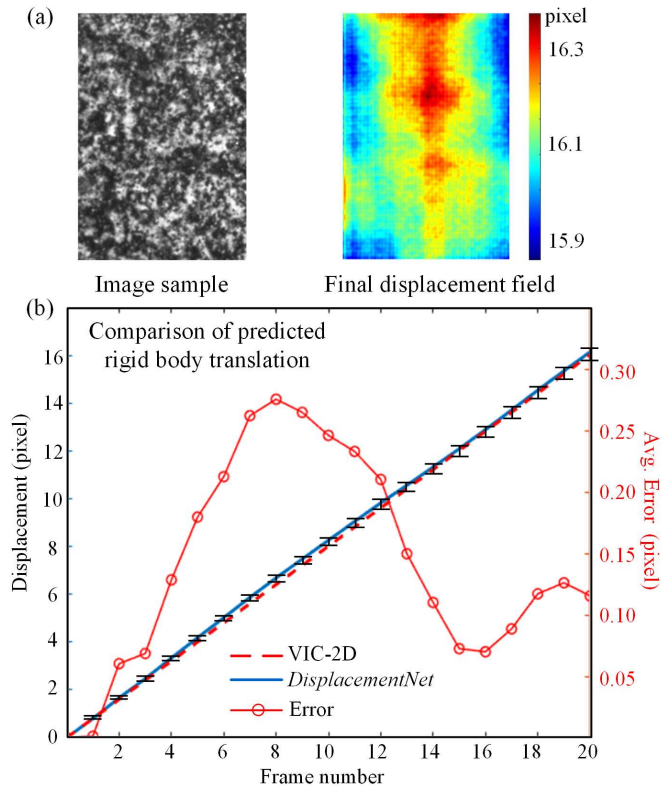

**Figure 12. (a) Predicted displacement field from *DisplacementNet* for rigid body translation; (b) comparison of predicted average displacement from *DisplacementNet* and VIC-2D. (The error bar indicates the maximum difference within the predicted displacement field for each frame.)**

## Experimental result on a tensile test sample of bronze

In this example, we test the performance of *Deep DIC* on real image sequences captured from tensile testing of a bronze sample. The tensile testing setup is shown in **Figure 13(a)** with a miniature universal material test system (μts, *Psylotech Inc.*, USA), a CMOS camera (MQ022MG-CM, *Ximea*, Germany) and lens with a fixed working distance of 139 mm and a magnification ratio of 0.3X (#58-428, *Edmund Optics*, USA). The test sample was made of Bronze 220 and prepared to the dog-bone shape by waterjet cutting according to the dimensions indicated in **Figure 13(b)**. The test piece was fixed at one end and pulled at the other end at a constant speed of 12 μm/s until fracture. One example of captured images during the test is shown in **Figure 13(c)**.

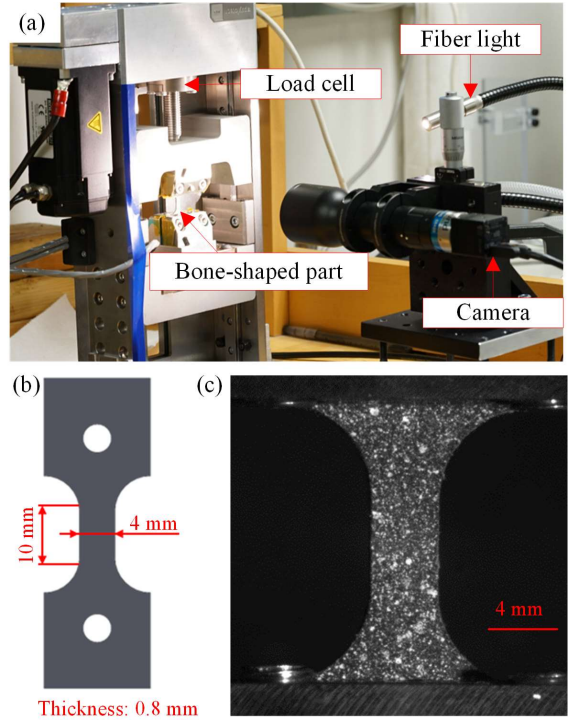

**Figure 13.** (a) Experimental setup; (b) bronze sample dimensions; (c) example of a captured image using the camera system.

We have collected a total of 189 images from the tensile test and fed them to both *Deep DIC* (*DisplacementNet* and *StrainNet*) and VIC-2D for comparison. We plot the predicted displacement and strain fields for three representative frames and compare their performance in **Figures 14** and **15**. The first frame at $T_1$ corresponds to the early stage when the deformation just starts. The second frame at $T_2$ is after yielding when localized deformation can be observed. The third frame at $T_3$ is towards the end of the experiment when the speckle patterns on the sample start to tear due to the crack initiation.

Our *Deep DIC* adaptively tracks the ROI, which starts at a size of 188 × 374 and ends at 178 × 434. The output resolution for both displacement and strain fields is the same as the input image size, so the local deformation variation can be better captured. For VIC-2D, we run the program using incremental correlation for a more stable prediction of large deformation. The subset and step sizes are set according to the suggested values by the software as 29 and 7, respectively. The output size from VIC-2D is about 1/7 of the original image size, thus giving an output size of 23 × 50, which stays the same for all predictions.

The VIC-2D results presented in **Figures 14 and 15** are interpolated to match the size of *Deep DIC* results for a direct comparison. Though VIC-2D results appear to be smoother in some predictions, it is not due to a better prediction accuracy but the interpolation operations.

The displacement prediction in the vertical direction is compared in **Figure 14**. Overall, we can observe very consistent and comparable displacement field predictions from *DisplacementNet* and VIC-2D. The absolute magnitude and the spatial distribution both match well. For time instances $T_1$ and $T_2$, VIC-2D results show clear quality fluctuations due to a large white spot in the speckle image as indicated in the figure inset. *DisplacemetNet* is less affected by the pattern variation since the inclusion of different quality speckle images in the training set.

The prediction of strain component $\varepsilon_{yy}$ is compared in **Figure 15**. Again, the overall magnitudes and spatial patterns match quite well between the two predictions. *StrainNet* shows larger noises in the small strain prediction at $T_1$, which is consistent with our noise floor level measurement. The circled areas in VIC-2D results at $T_1$ and $T_3$ show invalid predictions around edges and at the locations with very large deformation that the speckle patterns start to break, while *StrainNet* still produces reasonable results at these locations. This is one of the most significant advantages of *StrainNet*, where it is more robust than traditional DIC to handle different pattern variations even with edges and torn speckle patterns. These situations are more prevalent in testing polymer materials with extremely large strain, which will be further demonstrated and discussed in the next test case. In addition, *StrainNet* shows better spatial resolution to capture the localized strain concentration at the center of the sample in frame $T_3$, which matches the optical observation better as illustrated in the figure inset.

We would like to further comment on the achievable resolution and computation time for the *Deep DIC* and VIC-2D. The subset and step sizes in traditional DIC affect the output resolution and computation time. The subset size needs to be bigger enough to include sufficient pattern features for correlation, but also affects the spatial resolution. The step size directly controls the output size and affects the computation time by an inverse square relationship. That is to say, halving the step size will quadruple the calculation time. In *Deep DIC*, for both *DisplacementNet* and *StrainNet*, the prediction is performed on the pixel level, so the output image size will always equal the input image size. The computation time is scaled with the image input size, but not affected by the output resolution and speckle pattern quality. There are also fewer knob tuning settings once the model is fully trained. For this specific tensile test example with 189 frames, the calculation time with VIC-2D is about 27 seconds with a subset size of 29 and a step size of 7 (manually measured with a timer). *Deep DIC* takes only 2.35 seconds in total to calculate both the displacement and strain fields including image file loading and calculation, which corresponds to 12.5 milliseconds per frame.

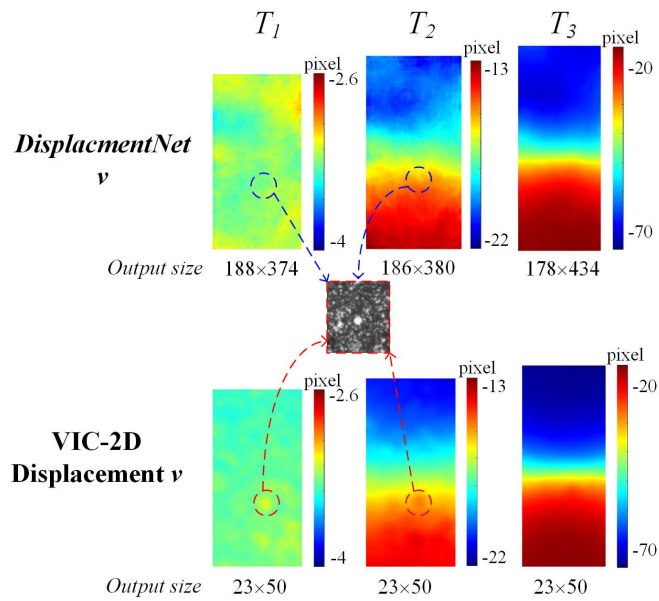

**Figure 14. Comparison of displacement prediction from *DisplacementNet* and VIC-2D for tensile testing on a bronze sample.**

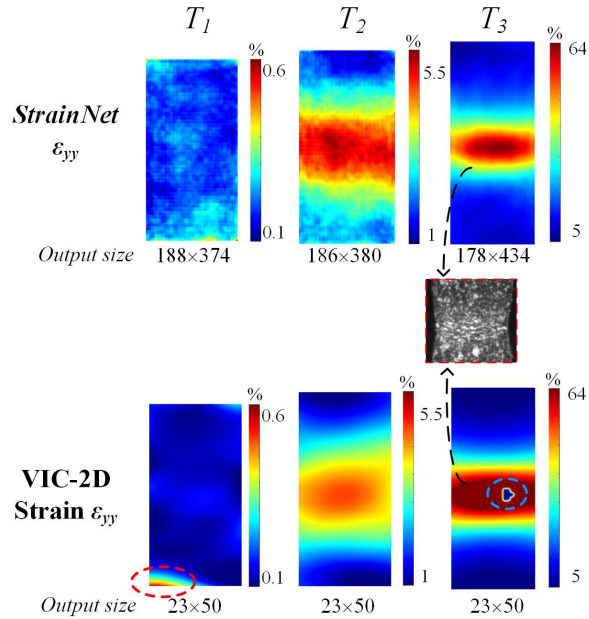

**Figure 15.** Comparison of strain prediction from *StrainNet* and VIC-2D for tensile testing on a bronze sample.

**<u>Validation on ultra-large strain measurement</u>**

In this example, we demonstrate the experimental results of strain prediction for tensile testing on an ultra-stretchable material. In this case, the accumulated strain can go up to more than 100%. A quasi-static tensile test was performed following ISO-8256 standard on a commercial-grade Polypropylene (PP) specimen, with dimensions shown in **Figure 16(a)**. The sample was stretched in the horizontal direction in the tensile test. The DIC measurement was performed using the Aramis 4M system (*GOM Metrology*, Germany). A total of 530 images were collected until the fracture of the sample. For GOM Correlate, the subset size and step size of 25 and 5 were used. The predicted strains from *Deep DIC* are compared with the results obtained from GOM Correlate. The x-direction strain $\varepsilon_{xx}$ is plotted and compared.

As shown in **Figure 16(b)** and **(c)**, the predicted strain fields are overlaid on the experimental images for a frame towards the end of the experiment. We notice very comparable strain distribution and absolute magnitudes as predicted by the two methods. The GOM Correlate has many invalid prediction zones, especially around the edges and cracks of the specimen, while *StrainNet* is quite robust for very large strain prediction and able to give a reasonable full-field prediction.

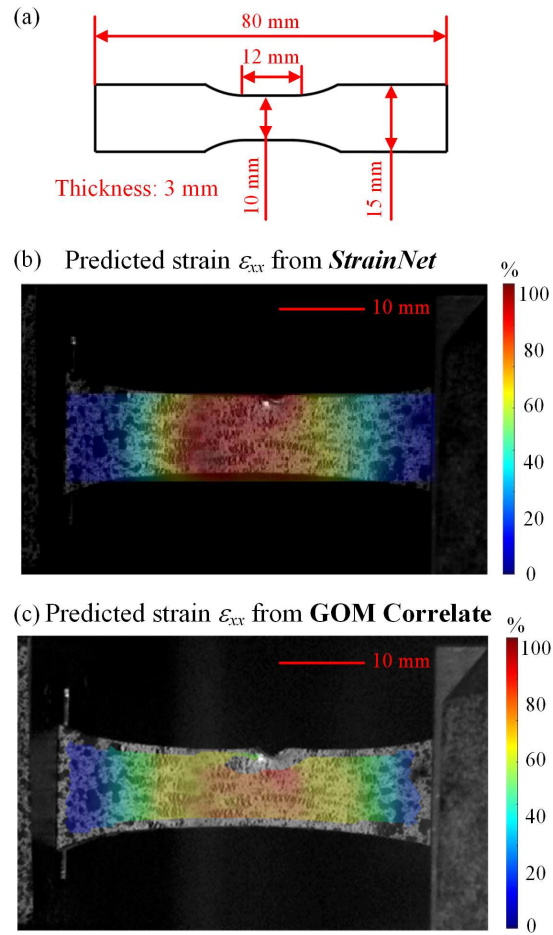

**Figure 16. (a) Polymer sample dimensions; Comparison of strain prediction between (b) *StrainNet* and (c) GOM Correlate for large deformation.**

A point-to-point comparison of predicted strain curves from *StrainNet* and GOM Correlate is shown in **Figure 17**. We pick six locations for comparison on the test sample that can be categorized into three groups depending on the level of deformation. Group 1 (points 1-3) has a cumulative strain large than 100% towards the end of the test. The predicted strain evolutions from *StrainNet* and GOM Correlate are in good consistency for strain up to 80%. After that, the GOM Correlate results become very unstable and frequently produce invalid values due to very localized deformation and tear of speckle patterns. The performance of traditional DIC is very sensitive to the quality of speckle patterns. *StrainNet* results are still reliable even with very large deformation. The second group (points 4 and 5) undergoes moderate deformation up to 70% strain, where the strain predictions from the two methods are highly consistent. The third group (point 6) is a very interesting group, which is located close to the clamping region. The strain predictions from the two methods are very different. Though no third strain measurement result is available to objectively quantify the prediction accuracy between the two, the strain curve predicted by *StrainNet* is more realistic. Since the sample is stretched slowly at a constant speed, point 6 will experience stretching first and a tendency of releasing motion due to the localized deformation observed in the middle region of the sample. The strain prediction of point 6 from GOM Correlate suggests two cycles of stretching and releasing, which is hard to find a plausible mechanics explanation. This additional strain fluctuation around frames 250-300 cannot be noticed in the strain curves from other points (1-5) either.

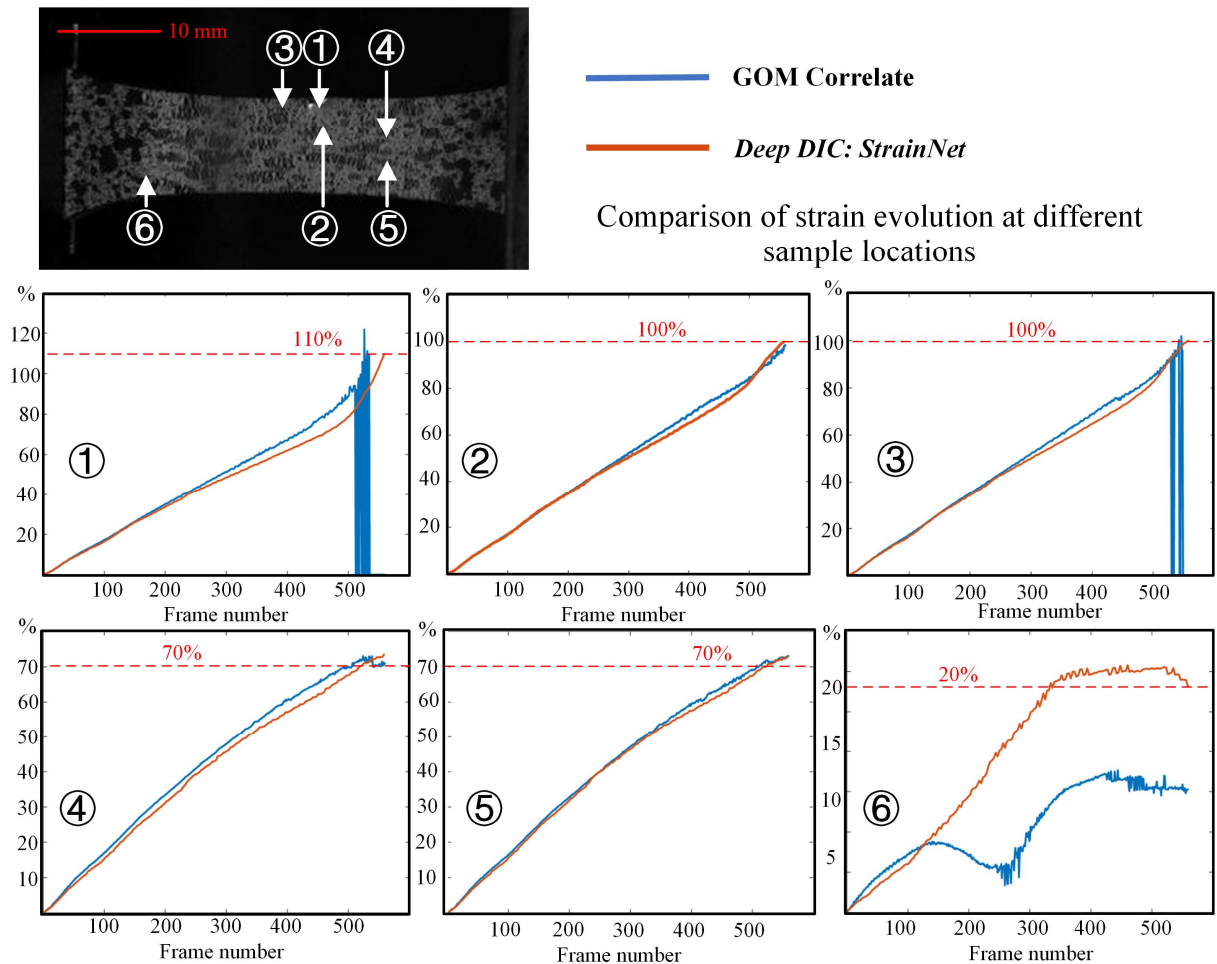

**Figure 17. A point-to-point comparison of strain curves for six selected points with *StrainNet* and GOM Correlate.**

For a total of 530 frames, *Deep DIC* starts with an ROI of 75 × 178 and ends with a final ROI of 72 × 317. The total running time, including image loading and calculation of displacement and strain fields, is 13.3 seconds, corresponding to 25.1 milliseconds per frame on average. For comparison, GOM Correlate takes more than 3 minutes in the calculation (with a subset size 25 and step size 5). The computational efficiency of traditional DIC is significantly influenced by speckle pattern quality as the speed dramatically drops when there is localized large deformation with deteriorated speckle patterns towards the latter frames. The computation speed of *Deep DIC* is quite stable and scaled with the image input size, but not affected by pattern quality.

## 4. Conclusion

In this paper, we develop a novel deep learning-based DIC method, *Deep DIC*, for end-to-end measurement of displacement and strain fields for material testing applications. Two CNNs, *DisplacementNet* and *StrainNet*, are developed to separately predict the displacement and strain fields from a pair of speckle images and to work collaboratively to adaptively update the ROI for tracking large deformation. To minimize the training cost, we develop a new method to generate a realistic and comprehensive training dataset including the reference and deformed speckle images, and the ground truths of predefined displacement and strain fields. The real-life performance of *Deep DIC*, including noise floor, rigid body motion tracking, strain measurement in tensile tests, etc., is systematically evaluated.

Two major innovations are achieved in this paper. (1) Compared with other deep learning-based DIC methods, *Deep DIC* utilizes a separate CNN, *StrainNet*, to achieve direct strain predictions from the image inputs, independent of the displacement measurement. The direct strain prediction from *StrainNet* avoids the large noises and errors induced by the discontinuity in the predicted displacement field. It preserves the high spatial resolution of strain prediction and does not require any post-filtering. In addition, *StrainNet* implicitly removes the influences of rigid body translation and motion from the strain calculation through its deep neural networks. (2) A new dataset generation method is developed to synthesize a realistic and comprehensive dataset, which critically affects the final performance of *Deep DIC*. To improve the model robustness, both high- and low-quality speckle patterns are generated to simulate the experimental conditions and image capture noises. Comprehensive and realistic deformation cases are included in the dataset, including rigid body translation and rotation, uniform stretch/compression, shear, and localized deformation formulated with 2D Gaussian functions.

Though *Deep DIC* is only trained on purely synthetic data, it achieves good performance on both simulated and experimental data. Compared with commercial DIC software, *Deep DIC* is able to (1) give highly consistent and comparable displacement and strain predictions for small and moderate deformation; (2) outperform commercial software in terms of robustness for strain predictions with large localized deformation and/or torn speckle patterns; (3) achieve more consistent and faster computation time down to the milliseconds level.

**Acknowledgment**

This work is supported by the start-up fund from McCormick School of Engineering, Northwestern University, Evanston, IL, USA.

**Appendix**

**A. Statistical analysis of the dataset**

We generate 40,150 pairs of specular images and the corresponding ground truths in total. The dataset is divided into a training set of 36,000, a validation set of 4,000, and a test set of 150. We perform a statistical analysis of the displacement and strain distributions in the training dataset to evaluate if the generated data give a good representation of a variety range of displacements and strains. The maximum displacement magnitude and its standard deviation within each sample are first calculated. Then we plot the statistical distributions of these two variables for all 36,000 samples of the training set in **Figure A-1(a)**. Similarly, the maximum strain magnitude and its standard deviation for all pixel values are calculated for each image. Their statistical distributions in the whole training set are plotted in **Figure A-1(b)**. The strain magnitude is taken as the equivalent strain.

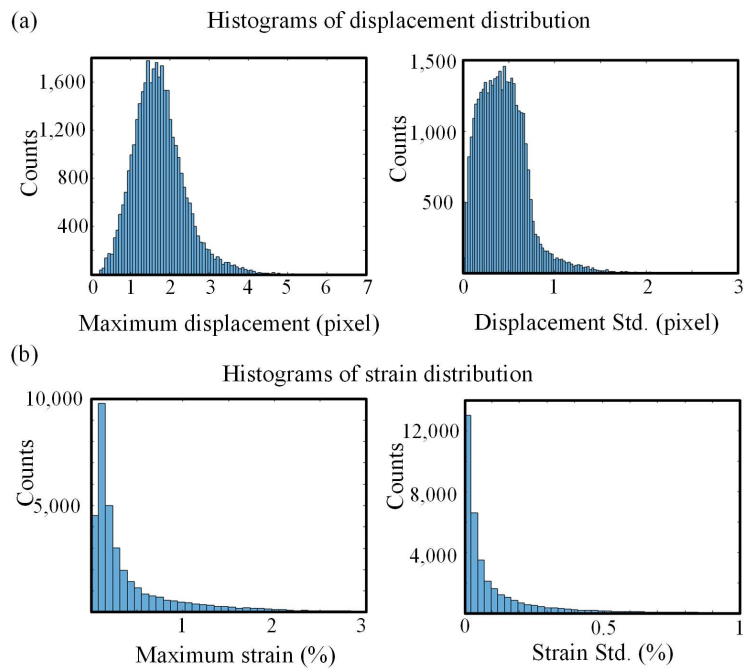

**Figure A-1.** Statistical analysis of the maximum and standard deviation for the (a) displacement and (b) strain distribution in the training set.

**B. Visualization of the feature maps**

    We further analyze the difference between *DisplacementNet* and *StrainNet* by visualizing and comparing the learned features maps in the two CNNs. In the encoder stage, each convolutional layer halves the feature map size but double its depth. We plot some representative features maps from the two CNNs for the first three convolution operations at different depths in **Figure B-1**. The first convolutional layer has a stack of 64 feature maps for both CNNs. The selective feature maps of depth 20 from *DisplacementNet* and of depth 2 from *StrainNet* are plotted for comparison. Both CNNs are extracting some low-level features, such as speckle boundaries, in the first convolutional layer. The feature maps of the two CNNs in the second convolutional layer are still topologically similar, but show a large point-to-point variance. Starting from the third layer, the differences between the two start to become obvious. For *DisplacementNet*, the feature map becomes localized, which indicates that the CNN tends to extract features from different regions, while *StrainNet* tends to have more uniformly distributed features. From a physics-based understanding, the strain calculation needs to remove the rigid body translation and rotation, while the displacement calculation only needs to calculate local correlation, so *StrainNet* requires more 'global' information than *DisplacementNet*.

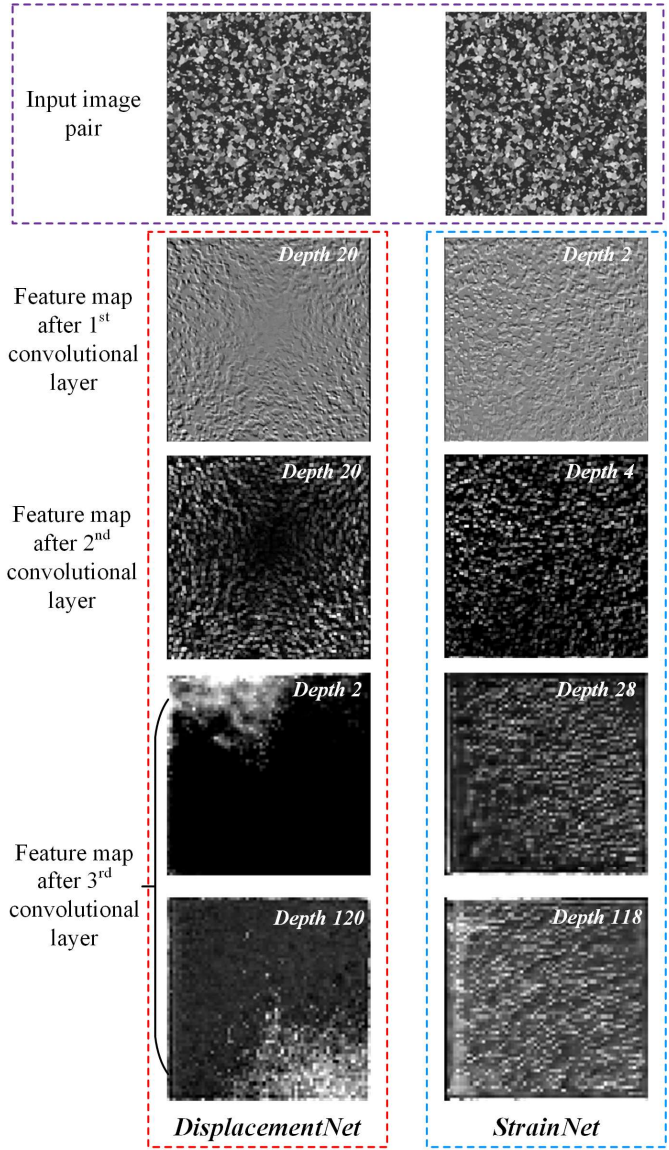

**Figure B-1.** Visualization of feature maps of the first three layers in *DisplacementNet* and *StrainNet*.

**List of Figures**

**List of Tables**